\RequirePackage{ifpdf}
\ifpdf 
\documentclass[pdftex]{sigma}
\else
\documentclass{sigma}
\fi

\usepackage{bm}

\newcommand{\g}{\gamma}
\newcommand{\lm}{\lambda}
\newcommand{\kp}{\varkappa}
\newcommand{\om}{\omega}
\newcommand{\oo}{\theta}
\newcommand{\ra}{\rangle}

\newcommand{\bg}{\boldsymbol{\gamma}}
\newcommand{\bs}{\boldsymbol{\sigma}}
\newcommand{\bu}{{\boldsymbol u}}
\newcommand{\bv}{{\boldsymbol v}}
\newcommand{\bt}{{\boldsymbol t}}

\newcommand{\bea}{{\cal{A}}}
\newcommand{\beb}{{\cal{B}}}
\newcommand{\bec}{{\cal{C}}}
\newcommand{\bed}{{\cal{D}}}
\newcommand{\tL}{{\widetilde L}}
\newcommand{\tR}{{\widetilde R}}
\newcommand{\tA}{{\widetilde {\cal A}}}

\begin{document}

\numberwithin{equation}{section}

\allowdisplaybreaks

\renewcommand{\PaperNumber}{013}

\FirstPageHeading

\renewcommand{\thefootnote}{$\star$}

\ShortArticleName{Relativistic Toda Chain with Boundary
Interaction at Root of Unity}

\ArticleName{Relativistic Toda Chain with Boundary Interaction\\
 at Root of Unity\footnote{This paper is a contribution
to the Vadim Kuznetsov Memorial Issue ``Integrable Systems and
Related Topics''. The full collection is available at
\href{http://www.emis.de/journals/SIGMA/kuznetsov.html}{http://www.emis.de/journals/SIGMA/kuznetsov.html}}}

\Author{Nikolai IORGOV~$^\dag$, Vladimir ROUBTSOV~$^{\ddag\S}$,
Vitaly SHADURA~$^\dag$ and Yuri TYKHYY~$^{\dag}$}
\AuthorNameForHeading{N.~Iorgov, V.~Roubtsov, V.~Shadura and
Yu.~Tykhyy}

\Address{$^\dag$~Bogolyubov Institute for Theoretical Physics, 14b
Metrolohichna Str.,  Kyiv, 03143 Ukraine}
\EmailD{\href{mailto:iorgov@bitp.kiev.ua}{iorgov@bitp.kiev.ua},
\href{mailto:shadura@bitp.kiev.ua}{shadura@bitp.kiev.ua},
\href{mailto:tykhyy@bitp.kiev.ua}{tykhyy@bitp.kiev.ua}}

\Address{$^\ddag$~LAREMA, D\'ept. de Math. Universit\'e d'Angers,
2 bd. Lavoisier, 49045, Angers, France}
\EmailD{\href{mailto:volodya@tonton.univ-angers.fr}{volodya@tonton.univ-angers.fr}}
\Address{$^\S$~ITEP, Moscow, 25 B. Cheremushkinskaja, 117259,
Moscow, Russia}

\ArticleDates{Received November 15, 2006, in f\/inal form January
03, 2007; Published online January 19, 2007}

\Abstract{We apply the Separation of Variables method to obtain
eigenvectors of commu\-ting Hamiltonians in the quantum relativistic
Toda chain at a root of unity  with boundary interaction.}


\Classification{81R12; 81R50}

\begin{flushright}
\emph{To the memory of Vadim Kuznetsov}
\end{flushright}

\section{Introduction}

During the last decade, a considerable progress in  application of
the Separation of Variables  method to a broad class of quantum
integrable models has been achieved. This progress was initiated
by the paper \cite{Skl2} of Sklyanin who has proposed (using the
$R$-matrix formalism) a recipe for a Separation of Variables in
the case of quantum Toda chain where the algebraic Bethe ansatz
fails. The next important step was done by Kharchev and Lebedev
\cite{KL1} who realized an iterative procedure of obtaining the
eigenfunctions of the $n$-particle open Toda chain by some
integral transformation from the eigenfunctions of the
$(n-1)$-particle open Toda chain. This iterative method was
applied later to a relativistic Toda chain \cite{KLS2}.

The Separation of Variables was found to be ef\/fective for the
derivation of the eigenfunctions of the commuting Hamiltonians of
the $n$-particle quantum Toda chain with   boundary
interaction~\cite{Kuzn,IorgShad} in the framework of Sklyanin's
approach to the integrable boundary problems \cite{SklB}.
 In the paper~\cite{IorgShad} the $n$-particle
eigenfunctions of the quantum Toda chain when the f\/irst and last
particles are exponentially interacting with the walls (the
two-boundary interaction) is constructed by means of an integral
transformation of the eigenfunctions for the Toda chain with
one-boundary interaction (the auxiliary problem). These
eigenfunctions, in turn, are constructed using the eigenfunctions
of the $n$-particle open Toda chain.

 Recently, a Separation of Variables was applied to the $N$-state  spin lattice
   model~-- the Baxter--Bazhanov--Stroganov (BBS) model \cite{GIPS, Iorgov}. In the
 $R$-matrix formulation  of this model the cyclic  $L$-operator \cite{BS, Kor}
 emerged. These operators are
  intertwined by the six-vertex $R$-matrix at a root of unity. The cyclic $L$-operator of the
BBS model reduces (at special values of parameters) to the
$L$-operator of the quantum Relativistic Toda Chain (RTC) at a
root of unity \cite{PS}.

The goal of this contribution is to f\/ind common eigenvectors of
commuting Hamiltonians of the RTC at a root of unity with boundary
interaction. We use for this the approach from \cite{IorgShad}.

Let us describe brief\/ly the structure of the paper. We start
with a short account of the model under consideration and remind
its properties based on the presentation in \cite{PS} (Section~2).
We recall the standard Sklyanin formalism introducing a chain of
$L$-operators satisfying some $RLL=LLR$ relations as well as its
``twisted'' version (\ref{l11}) such that the corresponding
(``twisted'') monodromy matrix satisf\/ies the ``Ref\/lection
Equations'' and some ``unitarity'' condition. We specify a
``boundary matrix'' solution $K_{+}(\lambda)$ of the \emph{dual
Reflection Equations} which are necessary for a description of a
two-boundary RTC. We close this section with a Hamiltonian
generating function to the two-boundary RTC.

We give the explicit formulas for the eigenvectors of the open RTC
in the Section 3 using the eigenvectors formulas for the open
Bazhanov--Stroganov quantum chain of which the open RTC is a
(specif\/ied) particular case \cite{BS}.

In Section 4 we consider the eigenvectors of the one-boundary RTC
specifying some particular choice of the boundary matrices. We
identify the coef\/f\/icients of the corresponding generating
function with the commuting one-boundary RTC Hamiltonians and
compute the common eigenvectors. The same questions are
treated in Section 5 for the two-boundary RTC using  the
Separation of Variables method. We write the Baxter equation for
separated variables explicitly~(\ref{true1}).

Section 6 contains an another proof of the commutativity of the
one-boundary RTC Hamiltonians of the Section~4. We show in fact
that they belong to a wide class of Hamiltonians which can be
expressed like a sort of Gelfand--Retakh quasideterminants. This
quasideterminants are (in our case) just Pl\"{u}cker coordinates
$\Delta_{s}\Delta_{0}^{-1} $ on $Gr_{n-1}(n)$.  The commutation
(and, hence the existence of the common spectrum) is guaranteed by
the results of \cite{ER} (see also \cite{Serg,BT}).

We f\/inish with a short discussion in the Section 7 and we leave
some tedious computations to the Appendix A.

We are dedicating this paper to the memory of Vadim Kuznetsov. His
important contributions to the subject of Separation of Variables
and in particularly to its applications in quantum integrable
models with boundary interaction \cite{Kuzn,Kuz2,KT,Kuz3} are
widely acknowledged in the Mathe\-matical Physics community. In
particular, it was Kuznetsov's paper \cite{KT} where the
Se\-pa\-ration of Variables was performed for the quantum
relativistic Toda chain and commuting Hamiltonians  of the quantum
RTC with integrable boundary interaction (the quantum RTC for
general classical root systems) were obtained.

\section{The formulation of the model}

In this section we recall brief\/ly the subject of the model
called the quantum Relativistic Toda Chain (RTC) chain at root of
unity \cite{PS}.

Let $\omega=e^{2\pi{\rm i}/N}$, $N\ge 2$. For each particle $k$,
$k=1,2,\ldots,n$, of the $n$-particle RTC there is a~corresponding
$N$-dimensional linear space (quantum space)~${\cal V}_k$ with the
basis $|\gamma\rangle_k$,$\gamma\in \mathbb{Z}_N$, and a pair of
operators $\{{\boldsymbol u}_k,{\boldsymbol v}_k\}$ acting on
${\cal V}_k$ by the formulas:
\begin{gather*}
{\boldsymbol v}_k|\gamma\rangle_k=v\omega^\gamma
|\gamma\rangle_k,\qquad {\boldsymbol u}_k|\gamma\rangle_k=u
|\gamma-1\rangle_k.
\end{gather*}
The space of quantum states of RTC with $n$ particles is ${\cal
V}={\cal V}_1\otimes{\cal V}_2\otimes\cdots\otimes{\cal V}_n$. We
extend the action of operators $\{{\boldsymbol u}_k,{\boldsymbol
v}_k\}$ to ${\cal V}$ def\/ining this action to be identical on
${\cal V}_s$, $s\ne k$. Thus we have the following commutation
relations
\begin{gather*}
{\boldsymbol u}_j {\boldsymbol u}_k={\boldsymbol u}_k {\boldsymbol
u}_j,\qquad {\boldsymbol v}_j {\boldsymbol v}_k={\boldsymbol v}_k
{\boldsymbol v}_j,\qquad {\boldsymbol u}_j {\boldsymbol v}_k=
\omega^{\delta_{j,k}} {\bv}_k{\boldsymbol u}_j.
\end{gather*}

Monodromy matrix for RTC with $n$ particles is
\begin{gather*}
T_n(\lm)=L_1(\lm)L_2(\lm)\cdots L_n(\lm)= \left( \begin{array}{ll}
A_n(\lm)& B_n(\lm)\\
C_n(\lm)& D_n(\lm) \end{array} \right),
\end{gather*}
where
\begin{gather}\label{Lop} L_k(\lm)=\left(
\begin{array}{cc}
1+\frac{\kp}{\lm}\bv_k& -\frac{1}{\lm}\bu_k\vspace{1mm}\\
\bu_k^{-1} \bv_k& 0 \end{array} \right),\qquad k=1,2,\ldots,n.
\end{gather} Thus the model depends on three parameters $u$, $v$
and $\kp$. It is possible to make this model to be inhomogeneous
by attaching the index $k$ to the parameters $u$, $v$ and $\kp$.
All the results presented in this paper can be directly
generalized to inhomogeneous case.

The straightforward calculation shows that the operators
$L_k(\lm)$ satisfy the following commutation relations:
 \begin{gather}\label{rll}
R(\lm,\mu)L_k^{(1)}(\lm)L_k^{(2)}(\mu)=L_k^{(2)}(\mu)L_k^{(1)}(\lm)R(\lm,\mu),\end{gather}
where the standard notations $L_k^{(1)}(\lm)=L_k(\lm)\otimes
\mathbb{I}$, $L_k^{(2)}(\mu)= \mathbb{I}\otimes  L_k(\mu)$ were
used and the $R$-matrix
$R(\lm,\mu)$ has the form \begin{gather*} 
R(\lm,\mu)=
\left(\begin{array}{cccc} \lm-\om\mu&0&0&0 \\
0&\lm-\mu& (1-\om)\mu&0 \\
 0& (1-\om)\lm& \om(\lm-\mu)&0\\
  0& 0& 0& \lm-\om\mu \end{array}\right).\end{gather*}
 Using (\ref{rll}), we obtain the following relation for monodromy matrix
\[
R(\lm,\mu)T_n^{(1)}(\lm)T_n^{(2)}(\mu)=T_n^{(2)}(\mu)T_n^{(1)}(\lm)R(\lm,\mu).
\]

We will use the method \cite{SklB} of  Sklyanin for description of
RTC with integrable boundary interaction. The $L$-operator
(\ref{Lop}) together with
\begin{gather}\label{l11} \tL_k(\lm)=\lm \sigma_x L_k^{-1}({\oo}/{\lm})\sigma_x=\left(
\begin{array}{cc}
\kp\lm+\om \oo\bv_k^{-1}& -\oo\bu_k^{-1}\vspace{1mm}\\
\ \lm \bv_k^{-1} \bu_k& 0 \end{array}
\right)=\om\oo\bv_k^{-1}\sigma_z L_k({\om\oo}/{\lm})^T
\sigma_z\end{gather} satisfy the following relations
\begin{gather*}
R(\lm,\mu)L_k^{(1)}(\lm)L_k^{(2)}(\mu)=L_k^{(2)}(\mu)L_k^{(1)}(\lm)R(\lm,\mu),\\
R(\lm,\mu)\tL_k^{(1)}(\lm)\tL_k^{(2)}(\mu)=
\tL_k^{(2)}(\mu)\tL_k^{(1)}(\lm)R(\lm,\mu),\\
\tL_k^{(2)}(\mu)\tR(\lm,\mu)L_k^{(1)}(\lm)=
L_k^{(1)}(\lm)\tR(\lm,\mu)\tL_k^{(2)}(\mu),\\
L_k^{(2)}(\mu)\tR(\lm,\mu)\tL_k^{(1)}(\lm)=
\tL_k^{(1)}(\lm)\tR(\lm,\mu)L_k^{(2)}(\mu),
\end{gather*}
where
\begin{gather*}
\tR(\lm,\mu)=\mu \sigma_x^{(2)}R(\lm,\oo/\mu)\sigma_x^{(2)}=
\left(\begin{array}{llll}\oo-\lm\mu&0&0&\om\oo-\oo\\0&\om\oo-\lm\mu&0&0
\\0&0&\om\oo-\lm\mu&0\\\om\lm\mu-\lm\mu&0&0&\om\oo-\om\lm\mu\end{array}\right),\\
\sigma_x=\left(\begin{array}{ll}0& 1\\1&0\end{array}\right),\qquad
\sigma_z=\left(\begin{array}{ll}1& 0\\0&-1\end{array}\right).
\end{gather*}
Here $\theta$ is an arbitrary parameter. It is important that
$\tR(\mu,\lm)=\tR(\lm,\mu)=P_{12}\tR(\lm,\mu) P_{12}$, where the
operator $P_{12}$ is the operator of permutation of two spaces.
Then the matrix
\[U(\lm)=\tL_{n}(\lm)\cdots\tL_{1}(\lm) K_{-}(\lm) L_{1}(\lm)\cdots L_{n}(\lm)=
\left(\begin{array}{ll}\bea(\lm)&\beb(\lm)\\\bec(\lm)&\bed(\lm)\end{array}\right)\]
satisf\/ies the {\it reflection equation}
\begin{gather}\label{ruru}R(\lm,\mu)U^{(1)}(\lm)\tR(\lm,\mu)U^{(2)}(\mu)=
U^{(2)}(\mu)\tR(\lm,\mu)U^{(1)}(\lm)R(\lm,\mu)\end{gather} if
$K_{-}(\lm)$ satisf\/ies the same equation:
\begin{gather}\label{five5}R(\lm,\mu)K_{-}^{(1)}(\lm)\tR(\lm,\mu)K_{-}^{(2)}(\mu)=
K_{-}^{(2)}(\mu)\tR(\lm,\mu)K_{-}^{(1)}(\lm)R(\lm,\mu).\end{gather}
The solution $K_{-}(\lm)$ with non-operator matrix elements is
\[K_{-}(\lm)=\left(\begin{array}{ll}\alpha_-(\frac{\oo}{\lm}-\lm)&\eta_-+\beta_-\frac{\oo}{\lm}\\
\eta_-+\beta_-\lm&\delta_-(\frac{\oo}{\lm}-\lm)\end{array}\right).
\] This solution
depends on four parameters $\alpha_-$, $\beta_-$, $\delta_-$,
$\eta_-$. It can be found directly solving~(\ref{five5}) or
adapting the formulas from \cite{VG} (see also \cite{KT}). The
matrix $U(\lm)$ satisf\/ies the so-called unitarity condition:
\begin{gather}\label{Uunit}
U(\om\oo/\lm)=\left(\begin{array}{ll}\bea(\om\oo/\lm)&\beb(\om\oo/\lm)\\
\bec(\om\oo/\lm)&\bed(\om\oo/\lm)\end{array}\right)\\
 =\frac{1}{\om(\oo-\lm^2)}
\left(\begin{array}{ll}(\lm^2-\om^2\oo)\bea(\lm)& \lm^2(1-\om)\beb(\lm)+(\om\oo-\lm^2)\bec(\lm) \\
\om(\om\oo-\lm^2)\beb(\lm)+\om \oo(1-\om)
\bec(\lm)&(\lm^2-\om^2\oo)\bed(\lm)\end{array}\right). \nonumber
\end{gather}

We have $\bea(\lm)\bea(\mu)=\bea(\mu)\bea(\lm)$ from the
ref\/lection equation (\ref{ruru}). It means that $\bea(\lm)$ is
generation function with respect to $\lm$ for commuting set of
operators. These operators are Hamiltonians for one-boundary RTC.

To describe a two-boundary RTC we need the {\it dual reflection
equation}:
\[R(\lm,\mu)K_+^{(2)}(\mu)Q(\lm,\mu)K_+^{(1)}(\lm)=
K_+^{(1)}(\lm)Q(\lm,\mu)K_+^{(2)}(\mu)R(\lm,\mu),\] where
\[Q(\lm,\mu)=\bigl( \big( \tR^{t_1}(\lm,\mu)\big)^{-1}\big)^{t_1}.
\]
Its solution is
\[K_{+}(\lm)=\left(\begin{array}{ll}\alpha_+(\frac{\om^2\oo}{\lm}-\lm)&\eta_++\beta_+\frac{\om\oo}{\lm}\\
\eta_+\om+\beta_+\lm&\delta_+(\frac{\om\oo}{\lm}-\om^{-1}\lm)\end{array}\right).\]
Then the generation function for the Hamiltonians of the
two-boundary RTC is given by $\bt(\lm)={\rm Tr}\,
(K_+(\lm)U(\lm))$ or explicitly:
\begin{gather*}
\bt(\lm)=\alpha_+(\oo\om^2/\lm-\lm)\bea(\lm)+
(\beta_+\lm+\eta_+\om)\beb(\lm)\\
\phantom{\bt(\lm)=}{} +(\beta_+ \oo\om /\lm+\eta_+)\bec(\lm)
+\delta_+(\om\oo/\lm-\om^{-1}\lm)\bed(\lm).
\end{gather*}

\section{Eigenvectors for the open RTC}

In order to give explicit formulas for the eigenvectors for RTC
we remind the def\/inition (see for example \cite{BB}) of $w_p(s)$
which is an analogue of $\Gamma_q$-function at root of unity. For
any point $p=(x,y)$ of Fermat curve $x^N+y^N=1$, we def\/ine
$w_p(s)$, $s\in \mathbb{Z}_N$, by
\begin{gather}\label{defw}
\frac{w_p(s)}{w_p(s-1)}=\frac{y}{1-x \omega^s},\qquad w_p(0)=1.
\end{gather}
Due to the Fermat curve condition the function $w_p(s)$ is cyclic:
$w_p(s+N)=w_p(s)$.

The explicit formulas for the eigenvectors of open RTC, that is
eigenvectors of $A_n(\lm)$, can be extracted from \cite{Iorgov},
where the formulas for the eigenvectors of open
Bazhanov--Stroganov quantum chain (which includes RTC) were
obtained. Here we give  the resulting formulas specialized to open
RTC with a little change of notations.

The eigenvectors of open RTC are given in terms of $w_p(s)$
def\/ined by the dif\/ferent points
$p_{m,s;m',s'}=(x_{m,s;m',s'},y_{m,s;m',s'})$,
$m,m'=1,2,\ldots,n;$
 $s=1,2,\ldots,m;$  $s'=1,2,\ldots,m',$   belonging to Fermat
curve $x^N+y^N=1$. The $x$-coordinates of these points are
def\/ined by unknown parameters (amplitudes) $v_{m,s}$ by
$x_{m,s;m',s'}=v_{m,s}/v_{m',s'}$. Then the corresponding
$y_{m,s;m',s'}$  are def\/ined up to a root of $1$. The parameters
$v_{m,s}$ are def\/ined  by the initial value $v_{11}:=v$ and the
recurrent relations
\begin{gather*}
v_{m1}v_{m2}\cdots
v_{mm}=v_{m-1,1}v_{m-1,2}\cdots v_{m-1,m-1} v, \\
\kp\prod\limits_{s\ne l} \frac{y_{m-1,s;m-1,l}}{y_{m-1,l;m-1,s}}
\, \frac{\prod\limits_{k=1}^m y_{m-1,l;m,k}}
{\prod\limits_{s=1}^{m-2} y_{m-2,s;m-1,l}}=1, \qquad
l=1,2,\ldots,m-1.
\end{gather*}
In the case of homogeneous RTC these relations can be solved
explicitly \cite{PS} (see also \cite{Iorgov}).

We will use the notation $|\boldsymbol{\gamma}_n\rangle \in{\cal
V}_1 \otimes\cdots\otimes{\cal V}_n$ for the eigenvectors of the
operator $A_n(\lambda)$ of the open RTC with $n$ particles. These
eigenvectors are labelled by $n$ parameters $\gamma_{n,s}\in
\mathbb{Z}_N$, $s=1,2,\ldots,n$, collected into a vector
\begin{gather*}
\boldsymbol{\gamma}_n=(\gamma_{n,1},\ldots,\gamma_{n,n})\in(\mathbb{Z}_N)^n.
\end{gather*}

The formula
\begin{gather*}
|\bg_n\ra=\sum_{\bg_{n-1}} \mu(\bg_{n-1}) Q(\bg_{n-1}|\bg_{n})
|\bg_{n-1}\ra\otimes |\sum_{k=1}^n \g_{n,k} - \sum_{l=1}^{n-1}
\g_{n-1,l}\ra_n,
\end{gather*} where
\begin{gather*}
\mu(\bg_{n-1})=\prod_{\genfrac{}{}{0pt}{}{j,l=1}{(j\ne l)}}^{n-1}
w_{p_{n-1,j;n-1,l}}(\g_{n-1,j}-\g_{n-1,l}-1), \\
Q(\bg_{n-1}|\bg_{n})=\frac{\om^{\g_{n-1,1}+\cdots+\g_{n-1,n-1}}}
{\prod\limits_{k=1}^n\prod\limits_{l=1}^{n-1}
w_{p_{n-1,l;n,k}}(\g_{n-1,l}-\g_{n,k}-1)},
\end{gather*} gives the eigenvectors $|\boldsymbol{\gamma}_n\rangle$ of $A_n(\lambda)$
from the eigenvectors $|\boldsymbol{\gamma}_{n-1}\rangle\in{\cal
V}_1 \otimes\cdots\otimes{\cal V}_{n-1}$ of $A_{n-1}(\lambda)$ and
basis vectors $|\gamma\rangle_n\in{\cal V}_{n}$. To f\/ind the
formula for $|\boldsymbol{\gamma}_{n-1}\rangle$ we can use the
same formula and so on. At the last step we need the eigenvectors
of $1$-particle RTC which are just basis vectors
$|\gamma_{1,1}\rangle_1\in{\cal V}_{1}$. The vectors $|\bg_n\ra$
obtained by the described procedure satisfy
\begin{gather*}
A_n(\lm)|\bg_n\ra=\prod_{k=1}^n\left(1+\frac{\kp
v_{n,k}\om^{\g_{n,k}}}\lm\right) |\bg_n\ra=
\prod_{k=1}^n\left(1-\frac{\lm_{n,k}}\lm\right) |\bg_n\ra, \\
B_n(\lm_{n,k})|\bg_n\ra=-\frac{h_k}{\lm_{n,k}}|\bg_n^{-k}\ra,\nonumber\\
B_n(\lm)|\bg_n\ra=-\frac{h_k}{\lm}\sum_{k=1}^n\prod_{s\ne k}
\left(\frac{\lm_{n,k}(\lm-\lm_{n,s})}{\lm(\lm_{n,k}-\lm_{n,s})}\right)
|\bg_n^{-k}\ra,
\end{gather*} where $\lm_{n,k}=-\kp v_{n,k}\om^{\g_{n,k}}$ and
$h_k=u\prod\limits_{l=1}^{n-1} y_{n-1,l;n,k}$. In the above
formulas the vector $|\bg_n^{-k}\ra$ means the vector
$|\g_{n1},\ldots,\g_{nn}\ra$ with $\g_{n,k}$ replaced by
$\g_{n,k}-1$.

We will omit in what follows the index $n$ in matrix elements of
monodromy matrix.

At $\lm=\om\mu$, the rank of $R(\lm,\mu)$ becomes $1$. Therefore
both sides of (\ref{rll}) become proportional to $R(\lm,\mu)$ with
the coef\/f\/icient being the so-called {\it quantum determinant}:
\begin{gather}\label{dabc} {\rm qdet}\;
T_n(\mu)=D(\om\mu)A(\mu)-B(\om\mu)C(\mu)= {\rm qdet}\;
L_1(\mu)\cdots {\rm qdet}\; L_n(\mu)= \frac{{\bf
V}}{(\om\mu)^n},\end{gather}
 where
\[{\bf V}=\prod_{k=1}^n \bv_k,\qquad {\bf V}|\bg_n\ra=\prod_{k=1}^n (v_{n,k}\om^{\g_{n,k}})\, |\bg_n\ra.\]
Now use (\ref{dabc}) for $\mu=\lm_{n,r}$:
\[D(\om \lm_{n,r})A(\lm_{n,r})-B(\om \lm_{n,r})C(\lm_{n,r})=
\frac{{\bf V}}{(\om \lm_{n,r})^n}\,.\]
 Acting by both sides on $|\bg_n\ra$ we get
\[-B(\om \lm_{n,r})C(\lm_{n,r})|\bg_n\ra=\frac{\prod\limits_{k=1}^n(v_{n,k}\om^{\g_{n,k}})}
{(\om \lm_{n,r})^n}|\bg_n\ra.\] Therefore
\[C(\lm_{n,r})|\bg_n\ra=\frac{\prod\limits_{k=1}^n(v_{n,k}\om^{\g_{n,k}}) }
{(\om \lm_{n,r})^{n-1}  h_{r}}|\bg_n^{+r}\ra.\] Since $C(\lm)$ is
a polynomial in $1/\lm$ of degree $n-1$ we can reconstruct the
action of $C(\lm)$:
\begin{gather*}
C(\lm)|\bg_n\ra=\sum_{r=1}^n\left(\prod_{m\ne r} \frac{\lm_{n,r}
(\lm-\lm_{n,m})} {\lm(\lm_{n,r}-\lm_{n,m})}\right)
C(\lm_{n,r})|\bg_n\ra .
\end{gather*}
Finally using (\ref{dabc}) after some calculation we get
\begin{gather*}
D(\lm)|\bg_n\ra=\frac{\om\kp}{\lm} \sum_{r=1}^n\sum_{k\ne
r}\Biggl[\frac{\prod\limits_m v_{n,m}\om^{\g_{n,m}}}{(-\kp
\om)^n}\left(\frac{v_{n,k}\om^{\g_{n,k}}}
{(v_{n,k}\om^{\g_{n,k}}-\om
v_{n,r}\om^{\g_{n,r}})(v_{n,r}\om^{\g_{n,r}}
-v_{n,k}\om^{\g_{n,k}})}\right)
\nonumber\\
\phantom{D(\lm)|\bg_n\ra=} {}\times\! \prod_{s\ne k,r}
\!\left(\frac{v_{n,k}\om^{\g_{n,k}} (\lm+\kp
v_{n,s}\om^{\g_{n,s}})}
{\lm(v_{n,k}\om^{\g_{n,k}}-v_{n,s}\om^{\g_{n,s}})(v_{n,r}\om^{\g_{n,r}}-v_{n,s}\om^{\g_{n,s}})}\right)\!
\left(\frac{h_{k}}{
h_{r}}|\bg_n^{+r-k}\ra-|\bg_n\ra\right)\Biggr].\!
\end{gather*}

\section{Eigenvectors for the one-boundary RTC}

We restrict our attention in what follows to the choice
$\beta_\pm=\delta_\pm=0$. In this case the boundary matrices
become
\[K_{-}(\lm)=\left(\begin{array}{ll}\alpha_-(\frac{\oo}{\lm}-\lm)&\eta_-\\
\eta_-&0\end{array}\right), \qquad
K_{+}(\lm)=\left(\begin{array}{ll}\alpha_+(\frac{\om^2\oo}{\lm}-\lm)&\eta_+\\
\eta_+\om& 0\end{array}\right).\] Due to (\ref{l11}) we have
\[\tL_n(\lm)\cdots\tL_1(\lm)=(\om\oo)^n {\bf V}^{-1}\sigma_z
T^t(\om\oo/\lm)\sigma_z=
(\om\oo)^n {\bf V}^{-1} \left(\begin{array}{ll}A(\om\oo/\lm)&-C(\om\oo/\lm)\\
-B(\om\oo/\lm)&D(\om\oo/\lm)\end{array}\right). \] Therefore
\begin{gather}
\bea(\lm)=(\om\oo)^n {\bf
V}^{-1}\!\left[\alpha_-\frac{\oo-\lm^2}{\lm}A(\om\oo/\lm)A(\lm)+
\eta_-(A(\om\oo/\lm)C(\lm)-C(\om\oo/\lm)A(\lm))\right]\!,\nonumber\\
\beb(\lm)=(\om\oo)^n {\bf
V}^{-1}\!\left[\alpha_-\frac{\oo-\lm^2}{\lm}A(\om\oo/\lm)B(\lm)+
\eta_-(A(\om\oo/\lm)D(\lm)-C(\om\oo/\lm)B(\lm))\right]\!,\nonumber\\
\label{calC} \bec(\lm)=(\om\oo)^n {\bf
V}^{-1}\!\left[-\alpha_-\frac{\oo\!-\!\lm^2}{\lm}B(\om\oo/\lm)A(\lm)+
\eta_-(-B(\om\oo/\lm)C(\lm)+D(\om\oo/\lm)A(\lm))\right]\!,\!\!\!\\
\bed(\lm)=(\om\oo)^n {\bf
V}^{-1}\!\left[-\alpha_-\frac{\oo-\lm^2}{\lm}B(\om\oo/\lm)B(\lm)+
\eta_-(-B(\om\oo/\lm)D(\lm)+D(\om\oo/\lm)B(\lm))\right]\!.\nonumber
\end{gather} From (\ref{Uunit})
it follows that $\tA(\lm)$ def\/ined by $\tA(\lm)={\lm
\bea(\lm)}/({\oo-\lm^2})$, or explicitly
\begin{gather}\label{aaac2} \tA(\lm)= (\om\oo)^n {\bf
V}^{-1}\left[\alpha_- A(\om\oo/\lm)A(\lm)+\frac{\eta_-
\lm}{(\oo-\lm^2)}(A(\om\oo/\lm)C(\lm)-C(\om\oo/\lm)A(\lm))\right],\end{gather}
satisf\/ies $\tA(\frac{\om\oo}{\lm})=\tA(\lm)$. It means that
$\tA(\lm)$ can be presented as
\begin{gather}\label{tAlm} \tA(\lm)=\alpha_- \kp^n  \left ( (\lm+\om\oo/\lm)^n +
(\lm+\om\oo/\lm)^{n-1} H^{\rm B}_1+\cdots+ (\lm+\om\oo/\lm) H^{\rm
B}_{n-1} +H^{\rm B}_{n}\right), \end{gather} where the set of
operators $H^{\rm B}_{1}, H^{\rm B}_2,\ldots,H^{\rm B}_{n}$ have
to be identif\/ied with the commuting set of Hamiltonians for the
one-boundary RTC. Explicitly, $H^{\rm B}_1$ is given by
(\ref{HBB1}) with $\eta_+=0$. Our problem is to f\/ind the common
eigenvectors for this set.

For $\bs=(\sigma_1,\ldots,\sigma_n)\in (\mathbb{Z_N})^n$, let
\begin{gather} \label{Psis}
\Psi_{\bs} =\sum_{\bg_n} Q(\bg_n,\bs)|\bg_n\ra, \end{gather} where
the sum includes all $N^n$ combinations of
$\bg_n=\{\g_{n,1},\ldots,\g_{n,n}\}$ and
\begin{gather}
Q(\bg_n,\bs)=\prod_{r=1}^n\om^{\g_{n,r}^2+\g_{n,r}} \prod_{k=1}^n
\prod_{r=1}^n\left(w_{p^{\rm B}_{r,k}}(\g_{n,r}-\sigma_k)
w_{\tilde p^{\rm B}_{r,k}}(\g_{n,r}+\sigma_k)\right)\nonumber \\
\phantom{Q(\bg_n,\bs)=}{} \times
\prod_{r<r'}\left(\frac{v_{n,r}\om^{\g_{n,r}}-v_{n,r'}\om^{\g_{n,r'}}}
{w_{\tilde p_{r,r'}}(\g_{n,r}+\g_{n,r'})} \right). \label{Qgs}
\end{gather}
Then, as it will be shown in the Appendix~A, we have
\begin{gather}
\tA(\lm)\Psi_{\bs}=\alpha_-\prod_{k=1}^n \left(\frac{(\lm+\kp
s_k\om^{\sigma_k})(\om\oo/\lm+\kp
s_k\om^{\sigma_k})}{s_k\om^{\sigma_k}}
\right)\Psi_{\bs}\nonumber\\
\phantom{\tA(\lm)\Psi_{\bs}}{} =\alpha_- (-\kp)^n \prod_{k=1}^n
\left(\frac{(\lm-\lm_k)(\om\oo/\lm-\lm_k)}{\lm_k}
\right)\Psi_{\bs} , \label{cAact}
\end{gather}
where $s_k$, $k=1,2,\ldots,n$, are some f\/ixed amplitudes which
will be def\/ined later. Also we have used a short notation
$\lm_k=-\kp s_k\om^{\sigma_k}$. This form of the eigenvalue
polynomial corresponds to~(\ref{tAlm}). We are using three types
of points $\tilde p_{r,r'}$, $p^{\rm B}_{r,k}$ and $\tilde p^{\rm
B}_{r,k}$  belonging to the Fermat curve $x^N+y^N=1$ in the
def\/inition of $Q(\bg_n,\bs)$. They are def\/ined by their
$x$-coordinates:
\[\tilde  x_{r,r'}=\kp^2 v_{n,r} v_{n,r'}/(\om\oo),\qquad
x^{\rm B}_{r,k}=v_{n,r}/s_{k},\qquad \tilde x^{\rm B}_{r,k}=\kp^2
v_{n,r} s_{k}/(\om\oo)
\]
and the equations
\begin{gather}\label{yyy}\frac{\eta_- v_{n,r}^2}{\kp^{n-2} \om^{n+1} \oo}
\prod_{r'\ne r}\tilde  y_{r,r'}=\alpha_- h_{r} \prod_{k=1}^n
y^{\rm B}_{r,k} \tilde y^{\rm B}_{r,k}, \qquad
r=1,2,\ldots,n.\end{gather} To solve these equations we have to
take the $N$-th degree of both parts of the equations (\ref{yyy})
and solve them with respect to elementary symmetric polynomials of
$s^N_k$, $k=1,2,\ldots,n$. This information is enough to f\/ind
the values of $s_k$ and the coordinates of the Fermat points
entering $Q(\bg_n,\bs)$.

Taking into account Laurent polynomial structure of ${\cal
C}(\lm)$ in $\lm$ it is clear that it is enough to know the action
formulas  for  ${\cal C}(\lm)$ on $\Psi_{\bs}$ in $2n+2$ points of
$\lm$ to reconstruct ${\cal C}(\lm)\Psi_{\bs}$. The following
formulas are valid:
\begin{gather} \label{actlmk} \bec(\lm_k)\Psi_{\bs}=
\eta_- f_k \Psi_{\bs^{+k}},\qquad \bec(\om\oo/\lm_k)\Psi_{\bs}=
\eta_-(\om\oo)^n f_k^{-1}\Psi_{\bs^{-k}},\\
\label{acttheta}
\bec(\pm\sqrt{\oo})\Psi_{\bs}=\eta_-(\pm\sqrt{\oo})^n \Psi_{\bs},
\end{gather}
where $\lm_k=-\kp s_k \om^{\sigma_k}$, $f_k=\prod\limits_{r=1}^n
\lm_k y^{\rm B}_{r,k}/\tilde y^{\rm B}_{r,k}$. The reconstructed
formula for ${\cal C}(\lm)$ is
\begin{gather}
{\cal C}(\lm)\Psi_{\bs}=\frac{\theta-\lm^2}{\lm} \sum_{k=1}^n
\prod_{r\ne k} \frac{\lm_k(\lm-\lm_r)(\om\theta-\lm\lm_r)}
{\lm(\lm_k-\lm_r)(\om\theta-\lm_k\lm_r)}\nonumber\\
\phantom{{\cal
C}(\lm)\Psi_{\bs}=}{}\times\frac{\lm_k}{\om\theta-\lm_k^2} \left(
\frac{\om\theta-\lm \lm_k}{\theta-\lm_k^2}\,
\bec(\lm_k)\Psi_{\bs}- \frac{\om\lm_k (\lm -
\lm_k)}{\om^2\theta-\lm_k^2}\,
\bec(\om\theta/\lm_k)\Psi_{\bs}\right)
\nonumber \\
\phantom{{\cal C}(\lm)\Psi_{\bs}=}{} +\frac{1}{2}
\sum_{\epsilon=\pm 1}\prod_{r=1}^n
\frac{(\lm-\lm_r)(\om\theta-\lm\lm_r)}
{\lm(\epsilon\sqrt\theta-\lm_r)(\om\epsilon\sqrt\theta-\lm_r)}
(1+\epsilon \lm/\sqrt\theta)
\bec(\epsilon\sqrt{\oo})\Psi_{\bs}.\label{Cfull}
\end{gather}

Here we give some heuristic explanation of formulas
(\ref{actlmk}). Let $\Psi_{\bs}$ be an eigenvector function of
$\bea(\lm)$ satisfying (\ref{cAact}). Then acting by both sides of
commutation relation
\begin{gather*}
(\om\mu-\mu)(\oo-\lm\mu)\bec(\lm)\bea(\mu)+(\mu-\lm)(\om\oo-\lm\mu)\bea(\lm)\bec(\mu)\\
\qquad\qquad +(\mu-\lm)(\om\lm\mu-\lm\mu)\beb(\lm)\bea(\mu)=
(\oo-\lm\mu)(\om\mu-\lm)\bec(\mu)\bea(\lm),
\end{gather*}
which follows from (\ref{ruru}), on $\Psi_{\bs}$  at $\mu=\lm_k$
we get
\[
\bea(\lm)\bec(\lm_k)\Psi_{\bs}=\frac{(\lm-\om\lm_k)(\oo/\lm-\lm_k)}{(\lm-\lm_k)(\oo\om/\lm-\lm_k)}
\bec(\lm_k)\bea(\lm)\Psi_{\bs}.
\]
Using (\ref{cAact}) we see that  $\bec(\lm_k)\Psi_{\bs}$ is an
eigenvector of ${\cal A}(\lm)$ with shifted zero, that is $\lm_k$
replaced by $\om\lm_k$. It means that $\bec(\lm_k)\Psi_{\bs}$ is
proportional to $\Psi_{\bs^{+k}}$. Clearly, this argumentation is
not suf\/f\/icient to prove the relations (\ref{actlmk}). The
derivation of these relations can be carried out adapting the
corresponding derivation from \cite{IorgShad}. To prove
(\ref{acttheta}) we use (\ref{calC}) at $\lm=\pm\sqrt{\oo}$ and
then (\ref{dabc}).

\section{Eigenvectors for the two-boundary RTC. \\
Separation of variables}

{}From (\ref{Uunit}) it follows
\[
\beb(\lm)=\frac{\bec(\lm)(\om-1)\oo+\bec(\frac{\om\oo}{\lm})(\oo-\lm^2)}{\om\oo-\lm^2},
\]
which gives
\begin{gather}
\bt(\lm)=\frac{\om^2\oo-\lm^2}{\lm}\alpha_+\bea(\lm)+\eta_+\bec(\lm)+\eta_+\om\beb(\lm)
\nonumber \\
\phantom{\bt(\lm)}{}
=\frac{(\om^2\oo-\lm^2)(\oo-\lm^2)}{\lm^2}\alpha_+\tA(\lm)+
\eta_+\frac{\bec(\lm)(\om^2\oo-\lm^2)+\om\bec(\frac{\om\oo}{\lm})(\oo-\lm^2)}{\om\oo-\lm^2}.
\label{tACC}
\end{gather}
We read of\/f the structure of $\bt(\lm)$ from this expression
\begin{gather*}
 \bt(\lm)=-({\om^2\theta^2}/{\lm^2}-\theta)(\lm^2-\theta)
\frac{\alpha_+ \alpha_- \kp^n}{\theta}\big ( (\lm+\om\oo/\lm)^n +
(\lm+\om\oo/\lm)^{n-1} H^{\rm BB}_1+\cdots
\nonumber\\
\phantom{\bt(\lm)=}{}    + (\lm+\om\oo/\lm) H^{\rm BB}_{n-1}
+H^{\rm BB}_{n}\big)
 +\eta_+\eta_- p_n(\lm), 
 \end{gather*}
where
\[ p_n(\lm)=(\lm+\om\theta/\lm)\theta^{(n-1)/2} \quad \mbox{if $n$ is odd},\qquad
p_n(\lm)=(1+\om)\theta^{n/2} \quad \mbox{if $n$ is even} \]
 and the set of operators $H^{\rm BB}_{1}, H^{\rm BB}_2,\ldots,
H^{\rm BB}_{n}$ should be identif\/ied with the commuting set of
Hamiltonians for the two-boundary RTC. Explicitly,
\begin{gather}\nonumber
H^{\rm BB}_1= \kp \sum_{k=1}^n \bigl( \bv_k + \om\theta\kp^{-2}
\bv_k^{-1}\bigr)- \sum_{k=1}^{n-1} \bigl(\bu_k\bu_{k+1}^{-1}
\bv_{k+1} +
\om \theta\kp^{-2} \bu_k \bu_{k+1}^{-1} \bv_{k}^{-1}\bigr)\\
\phantom{H^{\rm BB}_1=}{} -(\eta_-/\alpha_-) \bu_1^{-1} \bv_1
-(\eta_+/\alpha_+) \kp^{-1}\bv_n^{-1} \bu_n . \label{HBB1}
\end{gather}

Let $\Phi_{\bf E}$, ${\bf E}=(E_1,\ldots,E_n)$, be a common
eigenvector for the commuting set of operators $\{H^{\rm
BB}_{k}\}$:
\[
H^{\rm BB}_{k} \Phi_{\bf E}= E_k \Phi_{\bf E}\,
\]
and $t_{\bf E}(\lm)$ is corresponding eigenvalue polynomial for
$\bt(\lm)$. In accordance with a procedure of the Separation of
Variables we are looking for $\Phi_{\bf E}$ to be in the form
\[
\Phi_{\bf E}=\sum_{\bs} \prod_{i<j} \bigl((\lm_i-\lm_j) (\om\theta
-\lm_i\lm_j)\bigr) \prod_{k=1}^n q_k(\sigma_k,{\bf E})\,
\Psi_{\bs},
\]
where $\lm_k=-\kp s_k \om^{\sigma_k}$.

Using the  action formula (\ref{tACC}) for $\bt(\lm)$ on
$\Psi_{\bs}$ together with (\ref{cAact}) and (\ref{Cfull}) we get
Baxter equations for separated variables $q_k(\sigma_k,{\bf E})$:
\begin{gather} t_{\bf E}(\lm_k) q_k(\sigma_k,{\bf E})
=\eta_+\eta_-
\frac{(\oo-\lm_k^2)}{{(\om\oo-\om^{-2}\lm_k^2)}}\om^{1-2n}
f_k \,q_k(\sigma_k-1,{\bf E})\nonumber\\
\label{true1} \phantom{t_{\bf E}(\lm_k) q_k(\sigma_k,{\bf E})=}{}
 +\eta_+\eta_- \frac{(\om^2\oo-\lm_k^2)}{(\oo-\om\lm_k^2)}
\om^{n-1} \oo^n f_k^{-1}\, q_k(\sigma_k+1,{\bf E}),
\end{gather}
where $f_k=\prod\limits_{r=1}^n \lm_k y^{\rm B}_{r,k}/\tilde
y^{\rm B}_{r,k}$. The corresponding Baxter equation can be
considered (for each $k=1,2,\ldots, n$) as a homogeneous system of
linear equations for $N$ unknowns $q_k(\sigma,{\bf E})$,
$\sigma\in \mathbb{Z}_N$. The requirement of the existence of
non-trivial solutions for all $k=1,2,\ldots, n$ f\/ixes the common
spectrum for the commuting set of operators $\{H^{\rm BB}_{k}\}$.
Unfortunately, this problem can not be solved explicitly for
general $N$. But if we know the spectrum we can construct the
eigenvectors solving the systems of linear equations. We would
like to report about such a progress \cite{GIPS} in the case of
the periodic Baxter--Bazhanov--Stroganov model at $N=2$ which
contains in particular the Ising model.

\section[Another proof of commutativity of  $H^{\rm B}_{k}$]{Another proof of commutativity of
$\boldsymbol{H^{\rm B}_{k}}$}

In this section we describe an alternative approach to the proof
of commutativity of Hamiltonians using as an example the set of
the one-boundary RTC Hamiltonians  $H^{\rm B}_{k}$. This approach
is based on the results from \cite{ER}. Let us consider the
generation function
\begin{gather}
{\cal A}(\lm)=\alpha \prod_{k=1}^n \frac{(\lm+\tilde
\bv_k)(\theta/\lm+\tilde \bv_k)}{\tilde \bv_k} +\eta\sum_{k=1}^n
\tilde\bu^{-1}_r \tilde\bv_r \prod_{s\ne k}\frac{(\lm+\tilde
\bv_s)
(\theta/\lm+\tilde \bv_s)}{\tilde \bv_r-\tilde \bv_s}\nonumber\\
\phantom{{\cal A}(\lm)}{} =\alpha \prod_{k=1}^n (\Lambda+
\tilde\bv_k+\theta/\tilde\bv_k) +\eta\sum_{k=1}^n \tilde\bu^{-1}_k
\tilde{\bf V} \prod_{s\ne k} \frac{(\Lambda+
\tilde\bv_s+\theta/\tilde\bv_s)}{\tilde \bv_k-\tilde
\bv_s},\label{Aanother}
\end{gather}
where $\Lambda=\lm+\theta/\lm$, $\tilde{\bf V}=\tilde \bv_1 \tilde
\bv_2 \cdots \tilde \bv_n$. The set of operators $\tilde\bu_j$,
$\tilde\bv_j$ satisfy the relations
\begin{gather*}
\tilde{\boldsymbol u}_j \tilde{\boldsymbol u}_k=
\tilde{\boldsymbol u}_k \tilde{\boldsymbol u}_j,\qquad
\tilde{\boldsymbol v}_j \tilde{\boldsymbol v}_k= \tilde
{\boldsymbol v}_k \tilde{\boldsymbol v}_j,\qquad \tilde
{\boldsymbol u}_j \tilde{\boldsymbol v}_k= q^{\delta_{j,k}}
\tilde{\boldsymbol v}_k \tilde{\boldsymbol u}_j,
\end{gather*}
where $q$ is a non-zero complex number. Note that this expression
for ${\cal A}(\lm)$ coincides with (\ref{aaac2}) (see also
(\ref{tAact})) up to a redef\/inition of parameters.

Expanding ${\cal A}$ in $\Lambda$ we obtain
\begin{gather}\label{expanA}
{\cal A}=\alpha\, (\Lambda^n + H^{\rm B}_{1} \Lambda^{n-1} +\cdots
+ H^{\rm B}_{n-1} \Lambda+ H^{\rm B}_n).
\end{gather}
We will prove that the operators $H^{\rm B}_{k}$ obtained by such
expansion are pairwise commuting. Let us introduce new operators
$\Lambda_k=-\tilde\bv_k-\theta/\tilde\bv_k$. Taking into account
\[
\Lambda_k-\Lambda_s=(\tilde \bv_k-\tilde \bv_s)(\theta/(\tilde
\bv_k \tilde \bv_s)-1)\,,
\]
we get
\[
{\cal A}=\alpha \prod_{k=1}^n (\Lambda-\Lambda_k)
+\eta\sum_{k=1}^n \tilde Z_k \prod_{s\ne k}
\frac{\Lambda-\Lambda_s}{\Lambda_k-\Lambda_s},
\]
where
\[\tilde Z_k=\tilde\bu^{-1}_k \tilde \bv_k^n\prod_{s\ne k}\frac{ \tilde \bv_s^2}
{(\theta-\tilde \bv_k \tilde \bv_s)}.
\]
Finally using the interpolation formula for the polynomial
$\prod\limits_{k=1}^n (\Lambda-\Lambda_k)-\Lambda^n$ of degree
$n-1$:
\[
\prod_{k=1}^n (\Lambda-\Lambda_k)=\Lambda^n-\sum_{k=1}^n
\Lambda_k^n \prod_{s\ne k}
\frac{\Lambda-\Lambda_s}{\Lambda_k-\Lambda_s},
\]
we have
\begin{gather}\label{exprA}
{\cal A}= \alpha \Lambda^n +\alpha \sum_{k=1}^n Z_k \prod_{s\ne k}
\frac{\Lambda-\Lambda_s}{\Lambda_k-\Lambda_s},
\end{gather}
where $Z_k=\eta\tilde Z_k/\alpha-\Lambda_k^n$. It is easy to
verify that
\begin{gather}\label{comLZ}
[\Lambda_j,\Lambda_k]=0,\qquad [Z_j,Z_k]=0,\qquad
[Z_j,\Lambda_k]=0,\qquad j\ne k.
\end{gather}

\begin{theorem}
For the set of operators $\Lambda_k$, $Z_k$, $k=1,2,\ldots,n$,
satisfying \eqref{comLZ}, we have commutativity of operators
$H^{\rm B}_{k}$ def\/ined by expansion \eqref{expanA} of ${\cal
A}$ given by \eqref{exprA}.
\end{theorem}

\begin{proof}
Let $\boldsymbol{f}$ be the $n$ by $n+1$ matrix with the matrix
elements $f_{i,j}$, $1\le i \le n$, $0\le j \le n$, such that any
two of them belonging to dif\/ferent rows are commuting. Denote by
$\Delta_j$, $0\le j \le n$, the determinant of matrix
$\boldsymbol{f}$ with $j$-th column omitted. Then from \cite{ER}
it follows that the operators $H_k=\Delta_k \Delta_0^{-1}$,
$k=1,2,\ldots,n$, are pairwise commuting.

Let us f\/ix the matrix elements of $\boldsymbol{f}$ by the matrix
elements $f_{i,0}=Z_i$,  $f_{i,k}=\Lambda_i^{n-k}$, where $i$,
$k=1, 2,\ldots, n$. In this case we have
$\Delta_0=\prod\limits_{m<l} (\Lambda_m-\Lambda_l)$. Expanding
$\Delta_s$, $s\ge 1$,  with respect to f\/irst column we have
\[
\Delta_s=\sum_{k=1}^n Z_k (-1)^{k+1} \Delta_{k;0,s},
\]
where $\Delta_{k;0,s}$ is the determinant of matrix
$\boldsymbol{f}$ with $k$-th row and $0$, $s$-th columns omitted.
Therefore for the generating function for commuting $H_s=\Delta_s
\Delta_0^{-1}$ we get
\[
\sum_{s=1}^n (-1)^{s+1} H_s \Lambda^{n-s}= \sum_{k=1}^n Z_k
\left(\sum_{s=1}^n (-1)^{k+s} \Lambda^{n-s}  \Delta_{k;0,s}\right)
\Delta_0^{-1}.
\]
The expression in the brackets is the determinant expanded with
respect to $k$-th row of $\boldsymbol{f}$ with $0$-th column
omitted and all $\Lambda_k$ replaced by $\Lambda$. Hence this
expression is similar to $\Delta_0$ but with~$\Lambda_k$ replaced
by $\Lambda$. It gives
\[
\sum_{s=1}^n (-1)^{s+1} H_s \Lambda^{n-s}= \sum_{k=1}^n Z_k
\prod_{s\ne k} \frac{\Lambda-\Lambda_s}{\Lambda_k-\Lambda_s}.
\]
Comparing this formula with (\ref{expanA}) and (\ref{exprA}) we
get $H^{\rm B}_{k}=(-1)^{k+1} H_k$ and therefore the commutativity
of $H^{\rm B}_{k}$.
\end{proof}

Taking into account the discussion before the theorem we get that
the function (\ref{Aanother}) is a~generating function for the
commuting set of operators  $H^{\rm B}_{k}$.

\section{Discussion}
We had applied in this contribution the Separation of Variables to
obtain the commuting Hamiltonians eigenvectors of the quantum
relativistic Toda chain at a root of unity  with boundary
interaction. As we already have discussed in Introduction, the
$L$-operator of the relativistic quantum Toda chain at a root of
unity is a special case of the cyclic $L$-operator of the
Baxter--Bazhanov--Stroganov model (BBS). It was shown in the
papers (\cite{GIPS,Bugrij}), that the BBS model at $N=2$ coincides
with   the free fermion point of the generalized Ising model. We
plan to apply the Separation of Variables method, used in this
contribution, to the  $N=2$  BBS model with integrable boundary
conditions.

\appendix

\section{Derivation of the action formula for $\tA(\lm)$}

In this Appendix we will prove the formula
\begin{gather}
\tA(\lm)\Psi_{\bs} =\alpha_- (-\kp)^n \prod_{k=1}^n
\left(\frac{(\lm-\lm_k)(\om\oo/\lm-\lm_k)}{\lm_k}
\right)\Psi_{\bs}, \label{cAactAp}
\end{gather}
where the eigenvector $\Psi_{\bs}$ is def\/ined by (\ref{Psis}),
(\ref{Qgs}). In what follows we use the short notations
$\lm_{n,k}=-\kp v_{n,k}\om^{\gamma_{n,k}}$ and $\lm_k=-\kp
s_k\om^{\sigma_k}$.

To prove (\ref{cAactAp}) we use the expression (\ref{aaac2})
\begin{gather}\label{ac2Ap}\tA(\lm)= (\om\oo)^n
{\bf V}^{-1}\left[\alpha_- A(\om\oo/\lm)A(\lm)+\frac{\eta_-
\lm}{(\oo-\lm^2)}(A(\om\oo/\lm)C(\lm)-C(\om\oo/\lm)A(\lm))\right],
\end{gather}
and the action formulas
\begin{gather*}
A(\lm)|\bg_n\ra= \prod_{k=1}^n\left(1-\frac{\lm_{n,k}}\lm\right)
|\bg_n\ra,\qquad C(\lm)|\bg_n\ra=\sum_{k=1}^n\prod_{r\ne k}
\left(\frac{\lm_{n,k}(\lm-\lm_{n,r})}{\lm(\lm_{n,k}-\lm_{n,r})}\right)
C(\lm_{n,k})|\bg_n\ra,
\\ C(\lm_{n,k})|\bg_n\ra=\frac{\prod\limits_{m=1}^n \lm_{n,m}}
{(-\kp)^n h_k(\om\lm_{n,k})^{n-1}} |\bg_n^{+k}\ra.
\end{gather*}
We have
\begin{gather*}
(A(\om\oo/\lm)C(\lm)-C(\om\oo/\lm)A(\lm))|\bg_n\ra\\
\qquad{}=\frac{\theta-\lm^2}{\theta \lm} \sum_{k=1}^n \lm_{n,k}^n
\prod_{r\ne k} \left[\left(1-\frac{\lm \lm_{n,r}}{\om
\theta}\right)\left(1-\frac{\lm_{n,r}}{\lm}\right)
\frac{1}{\lm_{n,k}-\lm_{n,r}} \right] C(\lm_{n,k})|\bg_n\ra.
\end{gather*}
Therefore
\begin{gather}\label{tAact}
\tA(\lm)|\bg_n\ra= (\om\oo)^n \Biggl\{\alpha_- (-\kp)^n
\prod_{k=1}^n \frac{(1- \lm
\lm_{n,k}/(\om\theta))(1-\lm_{n,k}/\lm)}{\lm_{n,k}} |\bg_n\ra
\\
\nonumber \phantom{\tA(\lm)|\bg_n\ra=} {}+\frac{\eta_-}{\theta}
\sum_{k=1}^n \lm_{n,k}^n \prod_{r\ne k} \left[\left(1-\frac{\lm
\lm_{n,r}}{\om
\theta}\right)\!\left(1-\frac{\lm_{n,r}}{\lm}\right)\!
\frac{1}{\lm_{n,k}-\lm_{n,r}} \right]\!\frac{\om^{-1}}{
h_k(\om\lm_{n,k})^{n-1}} |\bg_n^{+k}\ra \Biggr\}.
\end{gather}
Acting $\tA(\lm)$ on $\Psi_{\bs}$ and shifting in an appropriate
way the summation we get
\begin{gather*}
\tA(\lm)\Psi_{\bs}= \sum_{\bg_n} Q(\bg_n,\bs)\tA(\lm) |\bg_n\ra
\\
\phantom{\tA(\lm)\Psi_{\bs}}{}= (\om\oo)^n \sum_{\bg_n}
Q(\bg_n,\bs) \Biggl\{\alpha_- (-\kp)^n \prod_{k=1}^n \frac{(1- \lm
\lm_{n,k}/(\om\theta))(1-\lm_{n,k}/\lm)}{\lm_{n,k}}
\\
\phantom{\tA(\lm)\Psi_{\bs}}{} +\frac{\eta_-}{\theta} \sum_{k=1}^n
\prod_{r\ne k}\! \left[\left(1-\frac{\lm \lm_{n,r}}{\om
\theta}\right)\!\left(1-\frac{\lm_{n,r}}{\lm}\right)
\frac{1}{\om^{-1}\lm_{n,k}-\lm_{n,r}} \right]\frac{\lm_{n,k} }{
h_k \om^{n+1}} \frac{Q(\bg_n^{-k},\bs)}{Q(\bg_n,\bs)}\Biggr\}
|\bg_n\ra.
\end{gather*}
Using (\ref{Qgs}) and (\ref{defw}) and then (\ref{yyy}) we obtain
\begin{gather*}
\frac{Q(\bg_n^{-k},\bs)}{Q(\bg_n,\bs)}=\frac{\kp^2
v_{n,k}^2}{\lm_{n,k}^2}
 \prod_{m=1}^n \frac{(\lm_m-\lm_{n,k})(1-\lm_m \lm_{n,k}/(\om\theta))}
{\lm_m y^{\rm B}_{k,m} \tilde y^{\rm B}_{k,m}}
\\
\phantom{\frac{Q(\bg_n^{-k},\bs)}{Q(\bg_n,\bs)}=}{} \times
\prod_{r\ne k}
\left(\frac{\om^{-1}\lm_{n,k}-\lm_{n,r}}{\lm_{n,k}-\lm_{n,r}}
\frac{\tilde y_{k,r}}{(1-\lm_{n,k}\lm_{n,r}/(\om\theta))}\right)
\\
\phantom{\frac{Q(\bg_n^{-k},\bs)}{Q(\bg_n,\bs)}}{}
=\frac{\alpha_-\theta \kp^n \om^{n+1} h_k}{\eta_- \lm_{n,k}^2}
 \prod_{m=1}^n \frac{(\lm_m-\lm_{n,k})(1-\lm_m \lm_{n,k}/(\om\theta))}
{\lm_m }
\\
\phantom{\frac{Q(\bg_n^{-k},\bs)}{Q(\bg_n,\bs)}=}{} \times
\prod_{r\ne k}
\left(\frac{\om^{-1}\lm_{n,k}-\lm_{n,r}}{\lm_{n,k}-\lm_{n,r}}
\frac{1}{(1-\lm_{n,k}\lm_{n,r}/(\om\theta))}\right).
\end{gather*}
Finally, taking into account the identity
\[
\prod_{k=1}^n (\Lambda-\Lambda_{n,k})+ \sum_{m=1}^n \prod_{r\ne m}
\frac{\Lambda-\Lambda_{n,r}}{\Lambda_{n,m}-\Lambda_{n,r}}
\prod_{k=1}^n (\Lambda_{n,m}-\Lambda_{k})=\prod_{k=1}^n
(\Lambda-\Lambda_{k})
\]
with \[ \Lambda=-\lambda-\om\theta/\lambda,\qquad
\Lambda_k=-\lambda_k-\om\theta/\lambda_k,\qquad
\Lambda_{n,k}=-\lambda_{n,k}-\om\theta/\lambda_{n,k}
\]
we get (\ref{cAactAp}).

\subsection*{Acknowledgements}

 The research presented
here is   supported  by the French--Ukrainian project ``Dnipro''
and the Ukrainian State Foundation for Fundamental Research. The
work of V.S. was also partially supported  by the SCOPES-project
IB7320-110848 of Swiss NSF. V.R. warmly acknowledged a partial
support of grants RFBR 06-02-17382 and NSh-8065.2006.2 as well as
the ANR-2005 (``Geometry and Integrability in Mathematical
Physics'') support.

\pdfbookmark[1]{References}{ref}
\LastPageEnding


\begin{thebibliography}{99}

\footnotesize\itemsep=0pt


\bibitem{Skl2}   Sklyanin E.,   Separation of variable. New trends,
{\it Prog. Theoret. Phys. Suppl.} {\bf 118} (1995), 35--60,
\href{http://arxiv.org/abs/solv-int/9504001}{\mbox{solv-int/9504001}}.


\bibitem{KL1}   Kharchev S.,   Lebedev D.,   Integral representation
for the eigenfunctions of quantum periodic Toda chain,
 {\it Lett. Math. Phys.} {\bf 50} (1999), 53--77,
\href{http://arxiv.org/abs/hep-th/9910265}{hep-th/9910265}.

\bibitem{KLS2}   Kharchev S.,   Lebedev D.,   Semenov-Tian-Shansky M.,
   Unitary representations of $U_{q}({sl}(2,R))$, the mo\-du\-lar double, and
the multiparticle $q$-deformed Toda chains, {\it
Comm.~Math.~Phys.} {\bf 225} (2002), 573--609,
\href{http://arxiv.org/abs/hep-th/0102180}{\mbox{hep-th/0102180}}.

\bibitem{Kuzn}  Kuznetsov V.,
Separation of variables for the $D_n$ type periodic Toda lattice,
{\it J. Phys. A: Math. Gen.} {\bf 30} (1997), 2127--2138,
\href{http://arxiv.org/abs/solv-int/9701009}{solv-int/9701009}.

\bibitem{IorgShad} Iorgov N., Shadura V.,  Wave functions of the Toda chain
with boundary interactions, {\it Theor. Math. Phys.} {\bf 142}
(2005), 289--305,
\href{http://arxiv.org/abs/nlin.SI/0411002}{nlin.SI/0411002}.

\bibitem{SklB}   Sklyanin E.,   Boundary conditions for
integrable quantum systems, {\it J. Phys. A: Math. Gen.} {\bf 21}
(1988), 2375--2389.

\bibitem{GIPS} von Gehlen G., Iorgov N., Pakuliak S.,
Shadura V.,   Baxter--Bazhanov--Stoganov model: separation of
variables and Baxter equation, {\it J. Phys. A: Math. Gen.} {\bf
39} (2006), 7257--7282,
\href{http://arxiv.org/abs/nlin.SI/0603028}{nlin.SI/0603028}.

\bibitem{Iorgov} Iorgov N.,      Eigenvectors of open Bazhanov--Stroganov quantum
chain, {\it SIGMA} {\bf 2} (2006), 019, 10 pages,
\href{http://arxiv.org/abs/nlin.SI/0602010}{nlin.SI/0602010}.


\bibitem{BS}
Bazhanov V.V., Stroganov Yu.G., Chiral Potts model as a descendant
of the six-vertex model, {\it J.~Statist. Phys.} {\bf 59} (1990),
799--817.

\bibitem{Kor}
Korepanov I.G., Hidden symmetries in the 6-vertex model of
statistical physics, {\it Zap. Nauchn. Sem. \mbox{S.-Peter\-burg.}
Otdel. Mat. Inst. Steklov. (POMI)} {\bf 215} (1994), 163--177
(English transl.: {\it J. Math. Sci. (New York)} {\bf 85} (1997),
1661--1670,
\href{http://arxiv.org/abs/hep-th/9410066}{hep-th/9410066}).

\bibitem{PS}
Pakuliak S., Sergeev S., Quantum relativistic Toda chain at root
of unity: isospectrality, modif\/ied $Q$-operator and functional
Bethe ansatz,  {\it Int. J. Math. Math. Sci.} {\bf 31} (2002),
513--554,
\href{http://arxiv.org/abs/nlin.SI/0205037}{nlin.SI/0205037}.

\bibitem{ER}
Enriquez B.,  Rubtsov V., Commuting families in skew f\/ields and
quantization of Beauville's f\/ibrations, {\it Duke Math. J.} {\bf
82} (2003), 197--219,
\href{http://arxiv.org/abs/math.AG/0112276}{math.AG/0112276}.

\bibitem{Serg} Sergeev S., Coef\/f\/icient matrices of a quantum discrete auxiliary
  linear problem, {\it Zap. Nauchn. Sem. POMI} {\bf 269} (2000),  no.~16, 292--307
  (in Russian).

\bibitem{BT} Babelon O.,  Talon M.,
Riemann surfaces, separation of variables and classical and
quantum integrability, {\it Phys. Lett.~A} {\bf 312} (2003),
71--77,
\href{http://arxiv.org/abs/hep-th/0209071}{hep-th/0209071}.


\bibitem{Kuz2} Kuznetsov V.B.,  Tsiganov A.V., Inf\/inite series of Lie
algebras and boundary conditions for integrable systems,  {\it J.
Sov. Math.} {\bf 59} (1992), 1085--1092.

\bibitem{KT}
  Kuznetsov V.B.,   Tsiganov A.V.,  Separation of variables for the quantum
relativistic Toda lattices,  Report 94-07, Mathematical Preprint
Series, University of Amsterdam, 1994,
\href{http://arxiv.org/abs/hep-th/9402111}{hep-th/9402111}.

\bibitem{Kuz3}   Kuznetsov V.B.,
Jorgensen  M.F.,   Christiansen P.L., New boundary conditions for
integrable lattices,  {\it J.~Phys.~A: Math. Gen.} {\bf 28}
(1995), 4639--4654,
\href{http://arxiv.org/abs/hep-th/9503168}{hep-th/9503168}.

\bibitem{VG} de Vega H.J., Gonzalez-Ruiz A.,
Boundary $K$-matrices for the XYZ, XXZ and XXX spin chains, {\it
J.~Phys.~A: Math. Gen.} {\bf 28} (1994),  6129--6141.

\bibitem{BB} Bazhanov V.V., Baxter R.J.,
Star-triangle relation for a three dimensional model, {\it J.
Statist. Phys.} {\bf 71} (1993), 839--864,
\href{http://arxiv.org/abs/hep-th/9212050}{hep-th/9212050}.


  \bibitem{Bugrij} Bugrij A.I., Iorgov N.Z., Shadura V.N.,
 Alternative method of calculating the eigenvalues of the transfer matrix of
the $\tau_2$  model for $N=2$,  {\it JETP Lett.} {\bf 119} (2005),
no.~2, 311--315.

\end{thebibliography}
\end{document}